\documentclass[11pt,leqno]{article}

\usepackage{amsmath,amsthm}
\usepackage {latexsym}
\usepackage{amssymb}

\newcommand{\bse}{\begin{equation}}

\newcommand{\bea}{\begin{eqnarray}}
\newcommand{\eea}{\end{eqnarray}}
\newcommand{\be}{\begin{equation}}
\newcommand{\ee}{\end{equation}}

\newcommand{\R}{{\mathbb R}}
\newcommand{\C}{{\mathbb C}}

\newcommand{\Z}{{\mathbb Z}}

\newcommand{\calg}{\,{\mathfrak g}}

\def\nn{\nonumber}
\def\calge1{\calg_{\vec{e_1}}}

\def\bm{\left( \begin{array}{cc}}
\def\endm{\end{array}\Big )}

\newtheorem{theorem}{Theorem}[section]
\newtheorem{lemma}[theorem]{Lemma}

\newtheorem{cor}[theorem]{Corollary}

\theoremstyle{remark}
\newtheorem{remark}[theorem]{Remark}

\numberwithin{equation}{section}

\begin{document}

\title{Dispersion for Schr\"odinger Operators with One-gap Periodic Potentials on $\R^1$}
 \author{ Kaihua Cai   \thanks{The author
feels deeply grateful to his advisor Wilhelm Schlag for his
guidance. Also the author thanks Irina Nenciu for reading a
preliminary version of this paper.}} \maketitle

\begin{abstract}
\noindent We prove $t^{- \frac 14}-$decay for the solutions of the
1-dim Schrodinger equation with a one-gap periodic potential as $t
\to +\infty $. Generically, one has $t^{- \frac 13}$-decay and this
decay is optimal. Our approach is to analyze the stationary phase in
the Schr\"odinger evolution as an integral operator.

\end{abstract}


\section{Introduction}

The dispersive property of the Schr\"odinger equation with
time-independent potentials

\be \frac 1i \partial_{t }\psi = -\Delta \psi + V(x) \psi
\label{eq:Sch0} \ee

\noindent has been extensively studied. It usually takes the
following form:

\be
 \| e^{iHt} P_{a.c.}(H) \varphi \|_\infty < C t ^{-\frac n2} \| \varphi
 \|_{L^1 (\R ^n)} \label{eq:decay0}
\ee

\noindent where $H= -\Delta + V(x) $, $n$ is the space dimension and
$P_{a.c.}(H) $ is the projection onto the absolutely continuous
spectral subspace. Most of the studies involve spatially decaying
potentials, for example \cite{R}, \cite{JK}, \cite{JSS}, \cite{Ya},
\cite{RS},\cite{GS}, \cite{Sc} and \cite{Go}.

In this article, we explore the dispersive property under the
assumption that $ V(x)$ is periodic, i.e.  $V(x)=V(x+2\omega )$,
and $n=1$. In this case, the differential equation

\begin{equation}
-y''(x) + V(x)y(x)=E y(x)  \label{eq:Hill}
\end{equation}

\noindent defined on the real line is known as Hill's equation. The
spectrum of the Schr\"odinger operator $-(d^2/dx^2) + V(x) $ acting
on $L^2(\R^1)$ is a union of intervals carrying purely absolutely
continuous spectrum. The absence of point spectrum and singular
spectrum suggests the dispersion of the solution. We will illustrate
this dispersive phenomenon for a special analytic potential, whose
spectrum is a union of two intervals (bands); namely, all gaps but
one are degenerate. It is known that one-gap potentials must be
elliptic functions (\cite{Ho}). With such a potential, we rewrite
Eq.\eqref{eq:Hill} as

\begin{equation}
y''(x) -2 \wp(x+\omega_3) y(x)=-E y(x) , \label{eq:Lame}
\end{equation}

\noindent where $\wp(z)$ is the Weierstrass elliptic function with
periods $2\omega_1, 2\omega_3$, satisfying the following
differential equations:

\bea \wp'(a)^2 = 4\wp^3(a)-g_2 \wp(a)-g_3  \label{eq:dif1}, \\
\wp''(a) = 6 \wp^2(a)-\frac {g_2}{2} \label{eq:dif2}. \eea

\noindent Hhere $g_2,\,g_3 $ are the invariants of $\wp(z)$ defined
by \eqref{eq:gg}. Eq.\eqref{eq:Lame}, known as Lam\'e's equation,
arises from the theory of the potential of an ellipsoid
(\cite{WW},\cite{Er}). We assume $\omega_1=\omega
>0, \,\, \omega_3= i \omega'$ and $\omega'
>0$ to guarantee that $\wp(x+\omega_3) $ is real-valued for $x \in \R$.
If we choose the potential in \eqref{eq:Lame} to be $n(n+1)\wp$,
instead of $2\wp$ ($n$ any positive integer), then the spectrum of
 Lam\'e's equation consists of  $n+1$ bands (\cite{MW}).

Eigenfunctions of \eqref{eq:Lame} are expressed in terms of the
Weierstrass $\sigma$-function and $\zeta$-function (\cite{KM},
\cite{GSS}) as follows:

\be  f_a(x)=\frac {\sigma (x+i\omega'+a)}{\sigma (x+i\omega')} e^
{-\zeta(a)x-\zeta(i\omega')a}, \label{eq:eigen} \ee

\noindent where the energy $$E=-\wp(a).$$

\eqref{eq:eigen} can be verified by noticing that (\cite{WW})

\be f'_a(x)= (\zeta(x+\omega_3 + a) -\zeta(x+\omega_3)  - \zeta( a)
 )f_a(x), \label{eq:fa'} \ee

\noindent and

$$ (\zeta(x+y)-\zeta(x)-\zeta(y) )^2 = \wp(x+y)+\wp(x)+\wp(y) .$$

Some basic properties of Weierstrass functions are listed in the
appendix. $f_a$ is periodic when $a$ is one of the half periods
$\omega_1,\omega_2=\omega_1+ \omega_3$ or $\omega_3$. $f_{-a}$ and
$f_a$ are the two Floquet-type solutions of \eqref{eq:Lame}. We
write

$$f_a(x)= m_a(x)e^{ik(a)x},$$

\noindent where

\be m_a(x)= \frac {\sigma(x+i\omega'+a)}{\sigma(x+i\omega')}
e^{-a\zeta(i\omega')-a \frac x \omega \zeta(\omega)} \label{eq:ma}
\ee

\noindent is periodic with period $2\omega$. Denote

$$\Sigma = [-\wp(\omega_1),-\wp(\omega_2)] \cup
[-\wp(\omega_3), +\infty ) ,$$

\noindent and the quasimomentum
$$k(a) = i\omega^{-1}(\omega \zeta(a)- a\zeta( \omega ))$$ is
real-valued for $E \in \Sigma$. $f_a $ is bounded when $ E \in
\Sigma $ and is unbounded otherwise, which implies that $\Sigma$ is
the spectrum of \eqref{eq:Lame} (\cite{MW}).

Our goal is to give a dispersive estimate similar to
\eqref{eq:decay0} for the following

\bea  \frac 1i \partial_{t }\psi(x,t) &=& -\frac {d^2}{dx^2}
\psi(x,t)
+ 2 \wp(x+\omega_3) \psi(x,t), \label{eq:Sch} \\
\psi(x,0)&=& \psi_0(x). \nn
 \eea

We assume that $\psi_0 \in L^1(\R) $ and denote the solution  at
time $t$ as $U(t)\psi_0$.

\begin{theorem} \label{thm:main1}

Generically, for almost all $\omega , \omega ' \in \R$, there
exists a constant $C> 0$ such that for $t>1$

 \be
 \| U(t)\psi_0 \|_{L^\infty(\R)} < C \, t ^{-\frac 13} \| \psi_0
 \|_{L^1 (\R)} .\label{eq:dec1}
\ee

\noindent Moreover, for all nonzero $\omega , \omega ' \in \R$,
there exists a constant $C> 0$ such that for $t>1$

\be
 \| U(t)\psi_0 \|_{L^\infty (\R)} < C \, t ^{-\frac 14} \| \psi_0
 \|_{L^1 (\R)} .\label{eq:dec2}
\ee

\eqref{eq:dec1}  is optimal in the sense that for any nonzero
$\omega, \omega' \in \R$, there exist constants $c>0$ and $ T
>0$, depending only on $\omega, \omega'$ such that for $t>T$

\be \sup_{\psi_0: \| \psi_0  \|_{L^1 (\R)}=1 } \| U(t)\psi_0
\|_{L^\infty(\R)} > c\, t ^{-\frac 13}. \label{eq:opti} \ee

\end{theorem}

We require $t >1$ to be large only to exclude $t \to 0$. The decay
rates $t ^{-\frac 13}$ and $t ^{-\frac 14}$ are different from $t
^{-\frac 12}$ in \eqref{eq:decay0} because phase function is
non-quadratic, which is a natural outcome of the periodic potential.
The decay factor $t ^{-\frac 13}$ as $t \to \infty$ has appeared in
the analysis of the Modified KdV equation (\cite{DZ}), where the
nonlinear phase of the main term is cubic. In our case, the analytic
phase function, roughly speaking, satisfies a cubic relation up to a
change of variables. This cubic relation comes from the differential
equations satisfied by the Weierstrass $\wp$ function . We denote $
P(x)$ to be the real-coefficient cubic polynomial

\be 2 x ^3 + \frac {6 \zeta(\omega)}{\omega} x^2 + \frac {g_2}2 x
+ g_3 -\frac {g_2 \zeta(\omega)}{2\omega}. \label{eq:cub}\ee

We shall prove \eqref{eq:dec1} under the assumption that

\be P(x ) \,\, \text{has no double root in} \,\,
(-\infty,\wp(\omega_3)]  . \label{eq:simple} \ee

If \eqref{eq:simple} does not hold, then we shall prove
\eqref{eq:dec2}. In this case, by Lemma~\ref{lemma:root}, $P(x)$ has
no root of degree $3$. Our proof implies that \eqref{eq:dec2} is
optimal in the sense stated in Theorem~\ref{thm:main1}. However, we
are unable to give an explicit example such that $P(x)$ does have a
double root in $(-\infty,\wp(\omega_3)]$.

Finally, we prove that assumption \eqref{eq:simple} holds for
almost all $\omega,\omega' \in \R$.

\section{Preliminaries}

It is known that $U(t)$ is an integral operator with kernel
$$K(t,x,x')=  \int_{\Sigma} e^{itE }  P_{a.c.}(E,x,x') dE .$$

\noindent Namely, $$ \psi(x,t) = \int_{\Sigma} \int_{\R}  e^{itE }
P_{a.c.}(E,x,x') \psi_0 (x') dx' dE  .$$

\noindent The absolutely continuous spectral projection is
$$P_{a.c.}(E,x,x')=\frac 1{2 \pi i } [(H-(E+i0))^{-1}(x,x')-(H-(E-i0))^{-1}(x,x')],
$$

\noindent and by definition $$(H-(E \pm i0))^{-1} =\lim_{\epsilon
\to 0^+} (H-(E \pm i \epsilon ))^{-1},$$

\noindent which can be expressed by $f_a$ and $f_{-a}$. Hence, we
obtain for $x> x'$

\begin{eqnarray}  K(t,x,x')& = & \int _\Sigma  e^{itE }
(f_{-a}(x')f_{a}(x)+f_{-a}(x)f_a(x') ) \frac{d E}{W(E)} \nonumber \\
& = & \int _\Sigma  e^{itE } ( e^{ik(a)(x-x')} m_{-a}(x')m_a(x)+
e^{-ik(a)(x-x')} m_{-a}(x)m_a(x') ) \frac {d E}{W(E)},
\label{eq:kernal}
\end{eqnarray}
\noindent where $W(E)=W(a)=W(f_a, f_{-a})=f_a f'_{-a}- f'_{a}f_{-a}
$, called the Wronskian of $f_a, f_{-a} $, is independent of $x$.

Because the spectral projection $P_{a.c.}$ is self-adjoint,
$P_{a.c.}(E,x,x') = \overline{P_{a.c.}(E,x',x)} $. Therefore, when
$x<x'$

\be K(t,x,x')= \int _\Sigma  e^{itE }
(\overline{f_{-a}}(x')\overline{f_{a}}(x)+\overline{f_{-a}}(x)\overline{f_a}(x')
) \frac{d E}{\overline{W}(E)}. \label{eq:kernal'}\ee

\noindent The proof of \eqref{eq:dec1} and \eqref{eq:dec2} shall be
reduced to proving

$$ \sup_{x,x'}|K(t,x,x')| < C t^ {-\frac 13} \,\,  \text{and } \,\, C t^ {-\frac 13} .$$

It follows from Eq.~\eqref{eq:fa'} that $$W(f_a, f_{-a})=
f_a(x)f_{-a}(x) ( \zeta(x+\omega_3 -a) +2 \zeta(a)- \zeta(x+\omega_3
+ a))
 .$$

\noindent Since $W(f_a, f_{-a}) $ is independent of $x$, we set
$x=0$ and obtain
$$W(f_a, f_{-a})=f_a(0)f_{-a}(0) ( \zeta(\omega_3 -a) +2 \zeta(a)-
\zeta(\omega_3 + a))  .$$

\noindent By the addition formula for Weierstrass functions
(\cite{Ak}, \S 15) $$ \zeta(u+v)- \zeta(u-v)-2 \zeta(v) =- \frac
{\wp '(v)}{\wp(u)-\wp(v)},$$

\noindent we have

\be W(E)= \frac {\sigma(i\omega ' +a) \sigma (i\omega '-a)
}{\sigma ^2(i\omega ')} \frac {\wp '(a)}{\wp(i\omega')-\wp (a)}.
\label{eq:WE} \ee  Hence,

\be  \frac {dE}{W(E)} = - \frac {\sigma^2(i\omega')(\wp
(i\omega')-\wp(a))}{ \sigma(i\omega'+a ) \sigma(i\omega'-a )
}da.\label{eq:WE1} \ee

\begin{remark}
 Because $\wp'(i\omega')=0 $ and zeroes of $\sigma $ are the lattice
 points
$\{ n_1 2\omega_1 + n_2 2\omega_3: n_1,n_2 \in Z \}$, all of which
are of degree $1$, it follows that $ \frac { \wp (i\omega')-\wp(a)}{
\sigma(i\omega'+a ) \sigma(i\omega'-a ) }$ is bounded and smooth
when  $  a \to i\omega'$.

 Also, it is clear that $\frac { \wp (i\omega')-\wp(a)}{
\sigma(i\omega'+a ) \sigma(i\omega'-a ) }= O(a^{-2}) $ when $a \to
0$, and that $\frac { \wp (i\omega')-\wp(a)}{ \sigma(i\omega'+a )
\sigma(i\omega'-a ) }$ and its $\partial_a$-derivatives are bounded
on $ [\omega_1, \omega_2]$, where $[\omega_1,\omega_2] $ denotes the
set
$$\{\lambda \omega_1 + (1- \lambda) \omega_2 : \lambda \in [0,1] \}
.$$

By \eqref{eq:ma}, $m_a$, $m_{-a}$ and their $\partial_a$-derivatives
are bounded uniformly for $a \in [\omega_1,\omega_2] \cup
[0,\omega_3] $, $x \in \R$.

\label{remark:ma} \end{remark}

Since $\Sigma$ is a union of two intervals, we shall decompose the
integral of $K(t,x,x')$ into two parts. Namely,

$$ K(t,x,x')= K_1(t,x,x')+ K_2(t,x,x'), $$
where $$K_1(t,x,x')= \int_{-\wp(\omega_1)}^{-\wp(\omega_2)}
e^{itE}P_{a.c.}(E,x,x') d E ,$$

 $$K_2(t,x,x')=
\int_{-\wp(\omega_3)}^{+\infty} e^{itE}P_{a.c.}(E,x,x') d E .
$$

Before we proceed to analyze $K_1(t,x,x')$ and $ K_2(t,x,x')$, we
prove two technical lemmas.

\begin{lemma} \label{lemma:stationary}

Let $ F(x) $ be a real-valued and smooth function on $(a,b)$,

\begin{enumerate}
\item Suppose $ |F'(x) | \geq \epsilon $, $ |F''(x) | \leq M $ for
all $x \in (a,b)$, then

$$ \Big | \int_a^b  e^{-it F(x)} \psi(x)dx \Big | \leq c \, \epsilon^{-1}  \, |t|^{-1}
 \Big [ |\psi(b)|+ \int_a^b (|\psi'(x)|+ |\psi(x)|)dx  \Big ], $$

\noindent where $c$ depends on $M$.

\item Suppose $k \geq 2 $, $k \in \Z $ and $ |F^{(k)}(x) | \geq
\epsilon$ for all $x \in (a,b)$, then
$$  \Big | \int_a^b  e^{-it F(x)} \psi(x)dx  \Big |
\leq c \, \epsilon^{-\frac 1k} \, |t|^{- \frac 1k} \Big [ |\psi(b)|+
\int_a^b|\psi'(x)|dx  \Big ],
$$

\noindent where $c$ depends on $k$.
\end{enumerate}

\end{lemma}

The first part of Lemma~\ref{lemma:stationary} follows from
integration by parts. The second part is proved in \cite{St} (p.
334).

\begin{lemma}
Let $e_j = \wp(\omega_j)$, $j=1,2,3$. Then $P(x)$ has a unique
simple root in $ [e_2, e_1] $, and $P(e_j)$, $j=1,2,3$, are nonzero.
Also $P(x)$ has no root of degree $3$ in $\R$. Moreover, $ -\frac
{\zeta(\omega)}{\omega} \in (e_3,e_2 )$. \label{lemma:root}
\end{lemma}

\begin{proof}

$P(x)=0$ if and only if $ 4 x^3 -g_2 x-g_3 = (6x^2- \frac
{g_2}2)(x+\frac {\zeta(\omega)}{\omega}) $. Denote $p_1(x)=4 x^3
-g_2 x-g_3  $ and $ p_2(x)=  (6x^2- \frac {g_2}2)(x+\frac
{\zeta(\omega)}{\omega})$. We shall examine the roots of $p_1(x)$
and $p_2(x)$ on the real line.

It follows from Eq.\eqref{eq:app3} that $p_1(x)=
4(x-e_1)(x-e_2)(x-e_3) $, where $e_j = \wp(\omega_j)$, $j=1,2,3$.
Because there is no quadratic term in $p_1(x)$, $e_1+e_2+e_3 =0$.
Since $e_3 < e_2 < e_1$, we have $e_3<0<e_1$.

Observe that $ \wp''(\omega_1) >0 $, $ \wp''(\omega_2) <0 $ and $
\wp''(\omega_3) >0 $, and by Eq~\eqref{eq:dif2}, we obtain

$$ \wp(\omega_2)^2 < \frac {g_2}{12} < \min \{ \wp(\omega_1)^2,\wp(\omega_3)^2 \} .$$

Now we shall prove $ \frac {\zeta(\omega)}{\omega} \in (-e_2, -e_3
)$. Indeed, let $y_1(x, E)$ and $y_2(x,E)$ be the solutions of
\eqref{eq:Lame} which satisfy

$$
y_1(0, E)=y'_2(0, E)=1,\qquad  y'_1(0, E)=y_2(0, E)=0.
$$

\noindent And we introduce the discriminant $ \Delta(E)=y_1(2\omega,
E)+ y'_2(2\omega, E)$.

Recall $a$ and $E$ are related by $E=-\wp(a)$, and as $E \to +
\infty$ on the real line, $a \to 0$ on the positive imaginary axis.
Therefore $ i \zeta(a)$ and $ k(a)$ go to $+\infty$ on the real line
when $E \to + \infty $ .

By Lemma 2.1 of \cite{GT}, $ \Delta(E) = 2\cos  k(a)$ and $k(E)=
k(a(E)) $ is the conformal map from the upper half plane to a slit
quarter plane $\Omega= \{ \Re z
>0,  \Im z > 0\} \backslash T   $, with the slit $T= \{ \frac
{\pi}{2\omega} + i y : 0< y \leq h \} $, where $h$ is some positive
real number. Moreover, $k(-e_1)=0 $ and $k(-e_2)=k(-e_3)= \frac
{\pi}{2\omega}$.

Denote $Q_0$ to be the pre-image of the tip $\frac {\pi}{2\omega} +
i h $ of the slit $T$ under the map $ k(E)$. Then $- \wp(\omega_2)<
Q_0 <  - \wp(\omega_3)$, and $k(E)$ sends $ [- \wp(\omega_2), Q_0]$
to $ [\frac {\pi}{2\omega}, \frac {\pi}{2\omega} + i h ] $, and $
[Q_0, - \wp(\omega_3)]$ to $ [ \frac {\pi}{2\omega} + i h, \frac
{\pi}{2\omega}] $ respectively. Thus when $E \in (- \wp(\omega_2),
Q_0)$, $\frac 1i \partial_E k(E) \geq 0 $. We observe that

$$ \partial_E k(E) = \frac 1{-\wp'(a)} \partial_a k(a) =
\frac i{\wp'(a)} \Big (\frac { \zeta( \omega )}{\omega}+ \wp(a)\Big
) ,
$$
 which implies that

$$ \frac {\wp(a) + \zeta (\omega)/\omega  }{ \wp'(a) } \geq 0.$$

\noindent Since $\wp'(a)>0 $ when $a \in (\omega_3,\omega_2)$, we
conclude that $ E= -\wp(a) \leq \zeta (\omega)/\omega $ for any $E
\in (- \wp(\omega_2), Q_0) $. Hence, $Q_0 \leq \zeta
(\omega)/\omega$. On the other hand, $\frac 1i \partial_E  k(E) \leq
0 $ when $E \in (Q_0, - \wp(\omega_3))$. Following the similar
argument, $Q_0 \geq \zeta (\omega)/\omega$. Therefore  $  \zeta
(\omega)/\omega= Q_0 \in ( - \wp(\omega_2), - \wp(\omega_3))$.

In fact, $k(E)$ maps $E= \frac {\zeta(\omega)}{\omega} $ to the tip
$\frac {\pi}{2\omega} + i h $ of the slit $T$ and $\Delta(E)$
reaches its minimum at $E=\frac {\zeta(\omega)}{\omega} $.

The three roots of $p_2(x)$ are $\pm \sqrt{ \frac {g_2}{12} } $ and
$  -\frac {\zeta(\omega)}{\omega} $. From the above analysis, we
have that $ \sqrt{ \frac {g_2}{12} } \in (e_2, e_1)$ and $ -\sqrt{
\frac {g_2}{12} }, -\frac {\zeta(\omega)}{\omega} \in (e_3,e_2)$,
which implies $p_2(e_1)
>0$ and $p_2(e_2) < 0$. Hence $P(x) $ has either one or three zeroes
in $(e_2,e_1)$ and clearly $P(e_j) $, $j=1,2,3$, are nonzero.

To verify that $P(x)$ has no root of degree $3$, we consider

$$ P'(x)= 6x^2 +12 \frac {\zeta(\omega)}{\omega} x + \frac {g_2}{2}.$$

\noindent The minimum of $ P'(x)$ is reached at $x =- \frac
{\zeta(\omega)}{\omega} \in ( e_3, e_2 )
 $ and is equal to $ \frac {g_2}{2} - 6(\frac {\zeta(\omega)}{\omega})^2 $.
 Notice that $ - \frac {\zeta(\omega)}{\omega} < e_2 < \sqrt {\frac
 {g_2}{12}}$ always holds.

If $- \frac {\zeta(\omega)}{\omega} > - \sqrt {\frac {g_2}{12}} $,
then $P'(x) >0 $ holds for all $x \in \R$. $P(x) $ has no root of
degree greater or equal to $2$.

If $- \frac {\zeta(\omega)}{\omega} = - \sqrt {\frac {g_2}{12}} $,
then $P'(x) $ has a double root $- \frac {\zeta(\omega)}{\omega} \in
(e_3,e_2) $. Since $P(x)$ has a root in $(e_2,e_1) $, we conclude
that $P(x)$ has no root of degree $3$.

If $- \frac {\zeta(\omega)}{\omega}< - \sqrt {\frac {g_2}{12}} $,
then $P'(x) $ has no double root. Hence $P(x)$ has no root of degree
$3$ on the whole real line.

If $P(x)$ has three zeroes in $ (e_2,e_1)$, then $P'(x)$ has two
roots in $(e_2,e_1)$, which is impossible because $ - \frac
{\zeta(\omega)}{\omega}  < e_2 $. Therefore, $P(x)$ has unique
simple root in $ (e_2,e_1)$.
\end{proof}

\section{Analysis of   $K_1(t,x,x') $ }

We first consider $K_1(t,x,x') $. We proceed by making the following
observation:

\begin{lemma} \label{lemma:reflect}

Let $b= 2\omega_2- a$ for $a \in [\omega_1,\omega_2]$. Write $W(a)=
W(f_a,f_{-a}) $. Then for $x,x' \in \R$

 \be \frac
{f_a(x')f_{-a}(x)}{W(a)} =- \frac {f_b(x)f_{-b}(x')}{W(b)}.
\label{eq:reflect} \ee
\end{lemma}

\begin{proof} It is clear that $
\wp(a)=\wp(b) $ and $ \wp'(a)=-\wp'(b) $. We prove
\eqref{eq:reflect} by direct calculation. By definition,

$$f_a(x')f_{-a}(x)= \frac {\sigma(x'+\omega_3+a)
\sigma(x+\omega_3-a)}
{\sigma(x+\omega_3)\sigma(x'+\omega_3)}e^{\zeta(a)(x-x')}. $$

\noindent By Eq.\eqref{eq:app1} and \eqref{eq:app2}, this equals

$$ \frac {\sigma(x'+\omega_3-b)
\sigma(x+\omega_3+b)}
{\sigma(x+\omega_3)\sigma(x'+\omega_3)}e^{\zeta(b)(x'-x)}
e^{4\eta_3 (\omega_3-b )} = f_b(x)f_{-b}(x')e^{4\eta_3 (\omega_3-b
)}  .$$

\noindent  Also by Eq.\eqref{eq:WE} and \eqref{eq:app2},

$$W(a)= \frac {\sigma(i\omega ' -b) \sigma (i\omega '+b) \exp {(4\eta_3 (\omega_3 -b))}
}{\sigma ^2(i\omega ')} \frac {-\wp'(b)}{\wp(\omega_3)-\wp(b)}
=-W(b) e^{4\eta_3(\omega_3 -b)}.
$$

\noindent  Combining them, \eqref{eq:reflect} follows.

\end{proof}

It follows from Lemma~\ref{lemma:reflect} that
$$ \int_{\omega_1}^{\omega_2} e^{-it\wp(a)} \frac {f_a(x')f_{-a}(x)}{W(a)} d\wp(a)
=  \int_{\omega_2}^{\omega_2 + i \omega'} e^{-it\wp(b)} \frac
{f_b(x)f_{-b}(x')}{W(b)} d\wp(b).$$

\noindent  Hence we have that for $x>x'$

\bea K_1(t,x,x') &=& \int _{ -\wp(\omega_1)}^{-\wp(\omega_2)}
e^{itE } (f_{-a}(x')f_a(x)+f_{-a}(x)f_a(x') )
\frac {dE}{W(E)} \nn \\
 &=& \int_{\omega_1}^{\omega_1 + i 2 \omega'}
e^{-it\wp(a)} \frac {f_a(x)f_{-a}(x')}{W(a)} d(-\wp(a)) \nn \\
&=& \int_{\omega_1}^{\omega_1 + i 2 \omega'} e^{-it\wp(a) +
i(x-x')k(a)} m_a(x)m_{-a}(x')\frac {-\wp'(a)da}{W(a)}.
\label{eq:part1} \eea

\noindent  Note $k(a)$ is real-valued and by Eq.\eqref{eq:kernal'},
we have that for $ x<x'$

$$ K_1(t,x,x') = \int_{\omega_1}^{\omega_1 + i 2 \omega'}
e^{-it\wp(a) +i (x-x')k(a)}
\overline{m_a}(x')\overline{m_{-a}}(x)\frac
{-\wp'(a)da}{\overline{W}(a)}.
$$

\noindent To simplify notation, we set $\tau= \frac {x-x'}{t} \in
\R$ and
$$F_\tau (a) = \wp(a)-i \tau (\zeta(a) -\frac a{ \omega}
\zeta(\omega)) .$$

\noindent Moreover, we write

\be K_1(t,x,x')= \int_{\omega_1}^{\omega_1 + i 2 \omega'}
e^{-itF_\tau(a)} \varphi(a,x,x') da, \label{eq:j1}\ee

\noindent  where $ \varphi(a,x,x') = m_a(x)m_{-a}(x')\frac
{-\wp'(a)}{W(a)}$ when $x>x'$, and $  \varphi(a,x,x')=
\overline{\varphi(a,x',x)}$ when $x<x'$. Without losing clarity,
$\varphi(a,x,x') $ will be written simply as $\varphi(a) $.

By Remark~\ref{remark:ma}, $\varphi(a,x,x') $ and its
$\partial_a$-derivatives are bounded uniformly for $a \in
[\omega_1, \omega_2] $ and $x,x' \in \R$. To apply
Lemma~\ref{lemma:stationary} to \eqref{eq:j1}, we analyze the
$\partial_a$-derivatives of $F_\tau(a)$. Our plan is to decompose
the integral in \eqref{eq:j1} into several regions and on each
region, Lemma~\ref{lemma:stationary} for some exponent $k$ will be
applied. We observe

\bea &&
\partial_a F_\tau(a)= \wp'(a)+ \tau i(\zeta(\omega)/\omega +
\wp(a)) ,\label{eq:c1} \\
&& \partial^2_a F_\tau(a)= \wp''(a)+ \tau i \wp'(a) ,\label{eq:c2} \\
&&
\partial^3_a F_\tau(a)=\partial^3_a \wp (a)+ \tau i \wp''(a). \label{eq:c3} \eea

\noindent Let

 $$ c_1=\min \{\zeta(\omega)/\omega + \wp(a): a \in [\omega_1, \omega_1+ 2i\omega']\}.$$ Then
$  c_1=\zeta(\omega)/\omega + \wp(\omega_2)$ and  by
Lemma~\ref{lemma:root}, $c_1 >0$. Also we denote

 \be M_1 =1+ \max
\{ |\wp'(a)|, |\wp''(a)|, \zeta(\omega)/\omega + \wp(a): a \in
[\omega_1,  \omega_1+ 2i\omega'] \} . \label{eq:M1}\ee

\noindent When $|\tau| > \frac {2M_1}{c_1}$,  we have for $a \in
[\omega_1,  \omega_1+ 2i\omega'] $

$$|\partial_a F_\tau (a)| > |\tau | c_1 - M_1 > \frac 12
|\tau | c_1 ,$$ and

$$|\partial^2_a F_\tau (a)| < M_1 ( |\tau |+ 1) . $$

Integrating by parts and recalling $\varphi(a)$ and its derivatives
are uniformly bounded, we obtain

\bea |K_1(t,x,x')| &=& \frac 1t \Big|  \int_{\omega_1}^{\omega_1 + i
2 \omega'} \frac {\varphi(a)}{\partial_a F_\tau (a) } d \, e^{-it
F_\tau (a)} \Big| \nn \\
  &\leq &   \frac { 4 \omega' \|\varphi(a)\|
_{L^\infty[\omega_1,\omega_2]} }   {t \tau c_1 } + \frac 1t \Big|
\int_{\omega_1}^{\omega_1 + i 2 \omega'} e^{-it F_\tau (a)} \Big(
\frac {\varphi ' (a)}{F'_\tau(a)} - \frac {\varphi (a) F''_\tau(a)
}{(F'_\tau(a))^2} \Big) da \Big|
  \label{eq:temp11} \\
 &\leq & C t^{-1}  .\nn  \eea

We now estimate $K_1(t,x,x')$ when $|\tau| \leq \frac {2M_1}{c_1}$.
Suppose both \eqref{eq:c1} and \eqref{eq:c2} vanish for $a=a_0 \in
[\omega_1,\omega_2]$ and $\tau = \tau_0 \in [-\frac {2M_1}{c_1},
\frac {2M_1}{c_1}]$. Then

$$ \wp'(a_0)^2= \wp''(a_0)( \frac {\zeta(\omega)}{\omega} + \wp(a_0)) .$$

\noindent By Eq.\eqref{eq:dif1} and \eqref{eq:dif2}, this is
equivalent to

$$ 2 \wp(a_0) ^3 + \frac {6 \zeta(\omega)}{\omega} \wp(a_0)^2 + \frac
{g_2}2 \wp(a_0) + g_3 -\frac {g_2 \zeta(\omega)}{2\omega}=0 .$$

\noindent Thus $\wp(a_0)$ is the simple root of $P(x)$ in
$[\wp(\omega_2),\wp(\omega_1)]$ and $\wp'(a_0) \neq 0$ by
Lemma~\ref{lemma:root}.

Observe that \eqref{eq:c1} and \eqref{eq:c2} also vanish when
$(a,\tau)= (2 \omega_2 - a_0,-\tau_0 )$. The analysis of $ (2
\omega_2 - a_0,-\tau_0 )$ is the same as that of $(a_0,\tau_0 )$ and
we will focus on $(a_0,\tau_0 )$.

Also we observe that $\partial^3_a F_\tau(a) $ vanishes at
$(a_0,\tau_0)$ if and only if

$$ \det \begin{pmatrix} \wp'(a_0 ) &  \frac {\zeta(\omega)}{\omega} + \wp(a_0 )
                               \\ \partial ^3_a \wp(a_0 ) & \wp''(a_0 ) \end{pmatrix}
=0;$$

\noindent namely,

$$ \partial_a \det \begin{pmatrix} \wp'(a ) &  \frac {\zeta(\omega)}{\omega} + \wp( a)
                               \\ \wp''(a ) & \wp'(a ) \end{pmatrix} _{a=a_0} =0  .$$

\noindent Since $\wp'(a_0) \neq 0 $, that $\partial^3_a F_{\tau_0}
(a_0)=0 $ is equivalent to the fact that $\wp(a_0)$ is a double root
of $P(x)$. By Lemma~\ref{lemma:root}, $P(x)$ has no double root in
$[\wp(\omega_2), \wp(\omega_1) ]$. Hence, $\partial^3_a F_{\tau_0}
(a_0) \neq 0 $ and there exists $\epsilon>0 $ such that

$$ \min \{ \Sigma_{j=1}^3 |\partial^{j}_aF_\tau (a)|:  a \in [\omega_1,\omega_1 +2i\omega'], \tau \in [-2M_1/c_1,
 2M_1/c_1] \} > \epsilon  > 0. $$

Let $ \chi_3(a,\tau) $ be a smooth function defined on $
[\omega_1, \omega_1 + 2i \omega'] \times [-\frac {2M_1}{c_1},
\frac{2M_1}{c_1}] $ such that $0 \leq \chi_3  \leq 1$, $
\chi_3(a,\tau) =1 $ when $ |\partial_a F_\tau (a)| + |\partial^2_a
F_\tau (a)| \leq \frac 13 \epsilon ,$ and $ \chi_3(a,\tau) =0 $
when $ |\partial_a F_\tau (a)| + |\partial^2_a F_\tau (a)| \geq
\frac 23 \epsilon$. Similarly, let $\chi_2(a,\tau)$ to be a smooth
function defined on $ [\omega_1, \omega_1 + 2i \omega '] \times
[-\frac {2M_1}{c_1}, \frac {2M_1}{c_1}] $, such that $0 \leq
\chi_2  \leq 1$, $ \chi_2(a,\tau) =1 $ when $|\partial^2_a F_\tau
(a)| \geq \frac 16 \epsilon $, and $ \chi_2(a,\tau) =0 $ when $
|\partial^2_a F_\tau (a)| \leq \frac 19 \epsilon$.


On the support of $\chi_3$, $|\partial^3_a F_\tau (a)| \geq \frac
13 \epsilon$. It follows from Lemma~\ref{lemma:stationary} that

\be \Big| \int_{\omega_1}^{\omega_1 + i 2 \omega'}e^{-it F_\tau (a)}
\chi_3(a,\tau) \varphi(a)da \Big | \leq C_3(\tau) t^{-\frac 13}.
\label{eq:temp3} \ee

On the support of $\chi_2 (1-\chi_3)$, $|\partial^2_a F_\tau (a)|
\geq \frac 19 \epsilon$. And similarly

\be \Big| \int_{\omega_1}^{\omega_1 + i 2 \omega'}e^{-it F_\tau (a)}
\chi_2(a,\tau) (1-\chi_3(a,\tau)) \varphi(a)da \Big | \leq C_2(\tau)
t^{-\frac 12}.\label{eq:temp4} \ee

On the support of $(1-\chi_2) (1-\chi_3)$, $|\partial^2_a F_\tau
(a)| \leq \frac 16 \epsilon$ and $|\partial_a F_\tau (a)| \geq
\frac 16 \epsilon$. Lemma~\ref{lemma:stationary} yields

\be \Big| \int_{\omega_1}^{\omega_1 + i 2 \omega'}e^{-it F_\tau (a)}
(1-\chi_2(a,\tau)) (1-\chi_3(a,\tau)) \varphi(a)da \Big | \leq
C_1(\tau) t^{-1}. \label{eq:temp5} \ee

\noindent Note $C_j(\tau), j=1,2,3$, are continuous functions of
$\tau \in [- 2M_1/ c_1,  2M_1/c_1 ] $. Let
$$C=
 \Sigma_{j=1}^3 \max \{ C_j(\tau): \tau \in [- 2M_1/ c_1,
2M_1 /c_1]\}.$$

\noindent Then $ |K_1(t,x,x')| \leq C t^{-\frac13}$ for large $t$,
because
$$ \chi_3 + \chi_2 (1-\chi_3)+ (1-\chi_2) (1-\chi_3)=1.$$

\noindent Consequently, we have proved that for large $t$

\be \sup_{x,x'} \big |K_1(t,x,x') \big | < C t^{-\frac 13},
\label{eq:fin1} \ee

\noindent where $C$ only depends on $\omega, \omega'$.

\section{Analysis of   $K_2(t,x,x') $ }

 We now consider $K_2(t,x,x')$. Let $b= 2\omega_3 - a$ for $a \in
(0,\omega_3)$, and the proof of Lemma~\ref{lemma:reflect} gives

$$\frac {f_a(x')f_{-a}(x)}{W(a)} =- \frac {f_b(x)f_{-b}(x')}{W(b)}   .$$

\noindent Then for $x>x'$

\bea K_2(t,x,x')&=& \int  _{-\wp(\omega_3)}^{+\infty}  e^{itE }
(f_{-a}(x')f_a(x)+f_{-a}(x)f_a(x') )
\frac {dE}{W(E)}\nn \\
&=& \int_{-\wp(\omega_3)}^{+\infty}
e^{-it\wp(a)} \frac {f_a(x)f_{-a}(x')}{W(a)} d(-\wp(a)) \nn \\
&=& \int_{0}^{i 2 \omega'} e^{-itF_\tau(a)} m_a(x)m_{-a}(x')\frac
{-\wp'(a)}{W(a)} da, \nn \eea

\noindent where $\tau=\frac {x-x'}{t}$. For $x<x'$, $K_2(t,x,x')$
can be written in a similar form. Therefore

$$ K_2(t,x,x') = \int_{0}^{i 2 \omega'} e^{-itF_\tau(a)}  \varphi (a,x,x') da,$$

\noindent where $\varphi (a,x,x')$ was defined in the previous
section.

\medskip
\textbf{Step 1.} The analysis of the nonlinear phase in
$K_2(t,x,x')$ is similar to that of $K_1(t,x,x')$. However, by
Remark~\ref{remark:ma}, $\varphi (a,x,x')$ in $K_2(t,x,x') $ is
unbounded when $a \to 0 $ and $a \to 2 i \omega'$, contrary to the
case of $K_1(t,x,x')$. Our strategy then is to change variables to
remove this singularity.

Define $ \lambda^2 = \wp(\omega_3) - \wp(a) $ such that $\lambda > 0
$ when $a \in (0, \omega_3)$ and $\lambda < 0 $ when $a \in
(\omega_3, 2 \omega_3)$. Then the map $ a \to \lambda $ is
one-to-one, onto and analytic from $ (0,2\omega_3)$ to $\R$. Note
that $\lambda(2i\omega' -a) = - \lambda(a) $ and the behavior of $
\lambda(a)$ as $ a \to  2i\omega'$ is the same as that when $ a \to
0$.

We claim that $\frac {\partial a}{\partial \lambda } = \frac
{2\lambda }{ -\wp'(a)}$ is never zero when $a \in (0,2\omega_3)$. In
fact, the claim is obvious for $a \neq \omega_3 $. When $a =
\omega_3 $,  by L'Hopital's Rule,

$$\frac {\partial a}{\partial \lambda
}(0)=\lim_{\lambda \to 0} \frac {2\lambda }{ -\wp'(a)} =
\lim_{\lambda \to 0} \frac 2{-\wp''(a) \frac {\partial a}{\partial
\lambda } }, $$

\noindent which implies that

$$ \Big |\frac {\partial a}{\partial
\lambda } (0) \Big | =  \sqrt {\frac 2 {\wp''(\omega_3)}}  >0.$$

Observe that when $\lambda \to \pm \infty$, $\lambda \varphi
(\lambda)=\lambda^3 \cdot O(1) $ and $|-\wp'(a)| = |\lambda|^3 +
O(\lambda^2) $. Hence, $\frac {\lambda \varphi (\lambda)}{-\wp'(a)}
$ and its $\lambda$-derivatives are bounded uniformly for $
x,x',\lambda  \in \R$.

After changing the variables, we obtain

\be \int_{0}^{i 2 \omega'} e^{-itF_\tau(a)} \varphi (a) da =
\int_{\R} e^{-itF_\tau(\lambda)} \varphi (\lambda) \frac {\partial
a}{\partial \lambda } d\lambda, \label{eq:main} \ee

\noindent where $ F_\tau(\lambda)= F_\tau(a(\lambda))$ and $\varphi
(\lambda) =\varphi (a(\lambda))  $.

We will decompose \eqref{eq:main} into different integral regions
and estimate them separately. Define $ \chi(\cdot) $ to be a smooth
function supported in $ (-2,2) $ such that $\chi(x)=1$ when $x \in
[-1,1]$, and let $M$ be a large number to be specified.

\medskip
\textbf{Step 2.} we claim

\be \Big | \int_{\R} e^{-itF_\tau(\lambda)} \varphi (\lambda) \frac
{2\lambda }{ -\wp'(a)} \chi(\lambda/M) d\lambda \Big |< C_M
\,t^{-\frac 14} \,\, \text{or} \,\,C_M t^{-\frac 13},
\label{eq:part2} \ee

\noindent depending on whether $P(x)$ has a double root in $ (-
\infty,\wp(\omega_3)]$ or not.

The proof of \eqref{eq:part2} will follow the lines of the proof of
\eqref{eq:fin1}. Recall that the map $\lambda \to a$ is one-to-one
from $\R$ onto $(0, 2\omega_3)$, and satisfies $ \lambda ^2 =
\wp(\omega_3) - \wp(a)$. Also, we observe

\bea && \partial_\lambda F_\tau (\lambda) =  \frac {\partial
a}{\partial \lambda } (\wp'(a)+ \tau i(\zeta(\omega)/\omega +
\wp(a))) ,\label{eq:temp12} \\
&& \partial^2_\lambda F_\tau (\lambda) =  \frac {\partial^2
a}{\partial \lambda^2 } (\wp'(a)+ \tau i(\zeta(\omega)/\omega +
\wp(a))) + \Big (\frac {\partial a}{\partial \lambda } \Big )^2
(\wp''(a)+ \tau i \wp'(a) ) . \label{eq:temp13}\eea

\noindent By Lemma~\ref{lemma:root}, $$\inf \{
|\zeta(\omega)/\omega + \wp(a) |: a \in (0, \omega_3) \} =
|\zeta(\omega)/\omega + \wp(\omega_3)| =c_2 >0 .$$

\noindent Denote $$M_2= \max \{ |\wp'(a)| + |\wp''(a)| +
|\zeta(\omega)/\omega + \wp(a) |:  \lambda(a) \in [-2M,2M] \} .$$

\noindent Since $\frac {\partial a}{\partial \lambda } $ is smooth
and never zero, there exist $c_3$ and $M_3$ such that  $0< c_3 <
|\frac {\partial a}{\partial \lambda }|< \sqrt{M_3}$  for all $
\lambda \in [-2M,2M] $. Moreover, suppose $ |\frac {\partial^2
a}{\partial \lambda ^2}|< M_3$ for $ \lambda \in [-2M,2M] $. Then
for $\lambda \in [-2M,2M] $ and $ |\tau| \geq 2 M_2/c_2 $

$$ |\partial_\lambda F_\tau (\lambda) | > \frac 12 c_2 c_3
|\tau| , \qquad \:   | \partial^2_\lambda F_\tau (\lambda)| < 2
M_3 M_2 (1+|\tau| ) .
$$

Integrating by parts, an argument similar to \eqref{eq:temp11}
shows that for $ |\tau| \geq 2 M_2/c_2 $

$$\Big | \int_{\R} e^{-itF_\tau(\lambda)} \varphi (\lambda) \frac
{2\lambda }{ -\wp'(a)} \chi(\lambda/M) d\lambda  \Big | < C_M
t^{-1}.$$

To prove \eqref{eq:part2} for $ |\tau| \leq 2 M_2/c_2 $, we first
suppose that $P(x)$ has no double root in $
(-\infty,\wp(\omega_3)]$. Since $ \frac {\partial
F_\tau(\lambda)}{\partial \lambda} = \frac {\partial a}{\partial
\lambda} \frac {\partial F_\tau(a)}{\partial a} $ and $ \frac
{\partial a}{\partial \lambda} \neq 0$ for $a \in (0,2\omega_3)$,
it follows from \eqref{eq:temp12} and \eqref{eq:temp13} that
$\frac {\partial F_\tau(\lambda)}{\partial \lambda} $ and $\frac
{\partial^2 F_\tau(\lambda)}{\partial \lambda ^2}$ vanish at
$(\lambda_0, \tau_0)$ if and only if \eqref{eq:c1} and
\eqref{eq:c2} vanish at $(a_0, \tau_0)$, where $ \lambda_0^2 =
\wp(\omega_3) - \wp(a_0) $. Therefore, the fact that $P(x)$ has no
double root implies that there exists $\epsilon  $ such that

$$ \min \Big \{ \sum_{j=1}^3 |\partial^j_\lambda F_\tau(\lambda)|: |\lambda| \leq 2 \lambda
  |\tau| < 2 M_2/c_2 \Big \} > \epsilon >0.$$

Just as in the case of $K_1(t,x,x')$, we define $\chi_2 (\lambda,
\tau), \chi_3(\lambda, \tau)$ on $[-2M,2M] \times [-2 M_2/c_2,2
M_2/c_2] $. Namely, $ \chi_2(\lambda, \tau)=1$ when $
|\partial^2_\lambda F_\tau(\lambda)| > \frac 16 \epsilon$, and $
\chi_2(\lambda, \tau)=0$ when $ |\partial^2_\lambda
F_\tau(\lambda)| < \frac 19 \epsilon$. $ \chi_3(\lambda, \tau)=1$
when $ |\partial_\lambda F_\tau(\lambda)| + |\partial^2_\lambda
F_\tau(\lambda)| < \frac 13 \epsilon$, and $ \chi_3(\lambda,
\tau)=0$ when $ |\partial_\lambda F_\tau(\lambda)| +
|\partial^2_\lambda F_\tau(\lambda)| > \frac 23 \epsilon$.

Decompose the integral in \eqref{eq:part2} according to
$$1 = \chi_3 + \chi_2(1-\chi_3) + (1- \chi_2)(1-\chi_3) .$$

\noindent The same arguments as that in  \eqref{eq:temp3},
\eqref{eq:temp4} and \eqref{eq:temp5} yield for $ |\tau| \leq 2
M_2/c_2 $

$$\Big | \int_{\R} e^{-itF_\tau(\lambda)} \varphi (\lambda) \frac
{2\lambda }{ -\wp'(a)} \chi(\lambda/M) d\lambda \Big |< C_M
t^{-\frac 13}.$$

In the case that $P(x)$ has a double root  $ \wp(a_0) \in
(-\infty,\wp(\omega_3)]$, there is $\tau_0 \in \R$, such that
$\partial^j_\lambda F_\tau(\lambda)$, $j=1,2,3$ vanish at
$(\lambda_0,\tau_0)$, where $\lambda_0 ^2 = \wp(\omega_3) -
\wp(a_0)$. The fact that $P(x)$ has no root of degree $3$ implies
that $\partial^4_\lambda F_{\tau_0}(\lambda_0) \neq 0 $.
Therefore, there exists $\epsilon $ such that

$$ \min \Big \{ \sum_{j=1}^4 |\partial^{j}_\lambda F_\tau (\lambda)|:
  \lambda \in [-2M,2M], \tau \in [-2M_2/c_2,
2M_2/c_2] \Big \} > 2 \epsilon  > 0. $$

Define smooth function $\chi_4(\lambda,\tau): [-2M,2M] \times
[-2M_2/c_2, 2M_2/c_2] \to [0,1]$, such that $\chi_4=1$ when
$\Sigma_{j=1}^3 |\partial^{j}_\lambda F_\tau (\lambda)| \leq
\epsilon $, and $\chi_4=0$ when $\Sigma_{j=1}^3
|\partial^{j}_\lambda F_\tau (\lambda)| \geq \frac 32 \epsilon $.
Hence on the support of $\chi_4 $, $ |\partial^4_\lambda F_\tau
(\lambda)| \geq \frac 12 \epsilon$. It follows from
Lemma~\ref{lemma:stationary} that

$$\Big | \int_{\R} e^{-itF_\tau(\lambda)} \varphi (\lambda) \frac
{2\lambda }{ -\wp'(a)} \chi(\lambda/M) \chi_4(\lambda, \tau)
d\lambda \Big |< C_4(\tau) t^{-\frac 14}.$$

We decompose the integral in \eqref{eq:part2} by using

$$ \chi_4+  (1-\chi_4)\chi_3 + \chi_2 (1-\chi_3)(1-\chi_4)+ (1-\chi_2) (1-\chi_3)(1-\chi_4)=1.$$

The analysis of the terms containing $(1-\chi_4)\chi_3, \chi_2
(1-\chi_3)(1-\chi_4)$ and $ (1-\chi_2) (1-\chi_3)(1-\chi_4)$ is
similar to  \eqref{eq:temp3},  \eqref{eq:temp4} and \eqref{eq:temp5}
respectively. Therefore, under the assumption that $P(x)$ has a
double root in $ (-\infty,\wp(\omega_3)]$, we have proved

$$\Big | \int_{\R} e^{-itF_\tau(\lambda)} \varphi (\lambda) \frac
{2\lambda }{ -\wp'(a)} \chi(\lambda/M) d\lambda \Big |< C_M
t^{-\frac 14}.$$

\medskip
\textbf{Step 3.} It now remains to estimate

$$\int_{\R} e^{-itF_\tau(\lambda)}
\varphi (\lambda) \frac {2\lambda }{ -\wp'(a)} (1-\chi(\lambda/M))
d\lambda, $$

\noindent which by definition equals

\be \lim_{N \to + \infty} \int_{\R} e^{-itF_\tau(\lambda)} \varphi
(\lambda) \frac {2\lambda }{ -\wp'(a)} (\chi(\lambda/N)-\chi(\lambda
/ M)) d\lambda .\label{eq:j2} \ee

\noindent Since $\frac {2\lambda \varphi (\lambda)}{ -\wp'(a)}$
For \eqref{eq:j2} are not integrable on the support of
$1-\chi(\lambda/M) $, Lemma~\ref{lemma:stationary} cannot be
applied to \eqref{eq:j2} directly.
 We shall explore the oscillation of the phase $e^{-itF_\tau(\lambda)}$
 and perform integration by parts to bound \eqref{eq:j2},
which requires us to exclude the zeroes of $\partial_\lambda
F_\tau(\lambda) $.

By definition, $\wp(a)= a^{-2} + \frac 1{20} g_2 a^2 + O(a^4),$
and $\wp(a) = \wp(i\omega') - \lambda ^2 $, hence

$$\lambda(a) = \frac ia + \alpha_1 a + O(a^3) \quad \text{as } a \to 0 , \,\, a \in (0,\omega_3), $$

\noindent which is an meromorphic function of $a$. It follows that
$\zeta(a) = -i \lambda + O(\lambda ^{-1}) $ as $a \to 0 ,\, a \in
(0,\omega_3)$. Consequently

 \bea F_\tau (\lambda) &=&  -\lambda ^2 - \tau \lambda+ \wp(i\omega') +
O( \frac \tau \lambda ),\quad \lambda \to \pm \infty ;
\label{eq:symp1} \\
   \partial_\lambda  F_\tau (\lambda)&=& -2\lambda - \tau +
O(\tau \lambda^{-2}) ,  \quad \lambda \to \pm \infty;
\label{eq:symp2} \\
 \partial^2_\lambda  F_\tau (\lambda)&=& -2 + O(\tau
\lambda^{-3}) ,  \quad \lambda \to \pm \infty. \label{eq:symp3}
\eea

\noindent We require $M$ large enough such that
$$M> 1+ \max \Big \{ |\lambda_0|: \frac {\partial
F_\tau(\lambda)}{\partial \lambda } , \frac {\partial^2
F_\tau(\lambda)}{\partial \lambda ^2} \,\, \text{both vanish at}
\, (\tau_0, \lambda_0)\Big \} .$$ Therefore, if $|\lambda|
> M$, $\frac {\partial F_\tau(\lambda)}{\partial \lambda }$
and $ \frac {\partial^2 F_\tau(\lambda)}{\partial \lambda ^2} $
cannot vanish at the same $(\tau, \lambda) $.

When $|\lambda | > M$ and $ |\lambda + \frac \tau 2|>1$, we claim

\be |\partial_\lambda F_\tau(\lambda)| > |\lambda + \frac {\tau}{2}|
- \frac 12 . \label{eq:grow} \ee

\noindent In fact, by \eqref{eq:symp2} $$|\partial_\lambda
F_\tau(\lambda)|
> 2|\lambda + \frac {\tau}{2}| - O(\tau \lambda ^{-2}). $$

\noindent We choose $M$ large enough such that $O(\tau \lambda
^{-2})< \frac {|\tau|} {100 |\lambda|}  $. If $|\lambda + \frac
{\tau}{2}| > \frac {|\tau|} {100 } $, \eqref{eq:grow} clearly
holds. If $|\lambda + \frac {\tau}{2}| \leq \frac {|\tau|} {100
}$, then $\frac {|\tau|} {100 |\lambda|} < \frac 12$ and
\eqref{eq:grow} also follows.

When $( - \frac \tau 2 -1,- \frac \tau 2 +1 )  $ is not contained
in $ (-M,M)$, by \eqref{eq:symp3}, $ |\partial^2_\lambda
F_\tau(\lambda)| > 1 $ for $ \lambda \in  ( - \frac \tau 2 -2,-
\frac \tau 2 +2 )$ as long as $M$ is large enough. By
Lemma~\ref{lemma:stationary}, we have

$$\int_{\R} e^{-itF_\tau(\lambda)} \varphi (\lambda) \frac {2\lambda }{
-\wp'(a)}  \chi(\lambda +\tau /2) d\lambda  < C t^{-\frac 12},
$$

\noindent where $\chi(x)= 1 $ when $|x|<1$, and $\chi(x)= 0 $ when
$|x|>2$.

To estimate \eqref{eq:j2}, we first consider the case when $ \frac
{|\tau|}{10} > M$ and estimate

$$ \int_\R e^{-itF_\tau(\lambda)}  \frac {2\lambda \varphi
(\lambda)}{ -\wp'(a)} (\chi(10 \lambda / |\tau|)-
\chi(\lambda/M)) d\lambda ,$$

\noindent which equals

\be \frac 1{it} \int_\R e^{-itF_\tau(\lambda)} \frac {d}{d \lambda}
\Big ( \frac {2\lambda \varphi (\lambda)}{ -\wp'(a)} \frac {\chi(10
\lambda / |\tau| )- \chi(\lambda/M)}{\partial_\lambda F(\lambda)}
\Big )d\lambda. \label{eq:temp2} \ee

\noindent It follows from Eq.\eqref{eq:symp2} that on the support
of $\chi(10 \lambda / |\tau|)- \chi(\lambda/M) $

$$|\partial_\lambda F_\tau(\lambda)|
>  |\tau| - 2|\lambda |- O(\tau \lambda^{-2}) > |\tau|/2  ,$$
and $$ |\partial^2_\lambda F_\tau(\lambda)| <  |\tau | /4,$$ as
long as $M$ is large enough. Hence

\be |\partial_\lambda  (\partial_\lambda F_\tau(\lambda))^{-1} | <
\frac 1 {|\tau|} . \label{eq:temp1} \ee

\noindent Since $(\chi(10 \lambda /| \tau|)- \chi(\lambda/M) )$,
$\frac {2\lambda \varphi (\lambda)}{ -\wp'(a)}  $ and their
$\lambda$-derivatives are uniformly bounded, we have

$$ \Big |\frac {d}{d \lambda} \Big( \frac {2\lambda \varphi (\lambda)}{
-\wp'(a)} \frac {\chi(10 \lambda / |\tau|)-
\chi(\lambda/M)}{\partial_\lambda F(\lambda)} \Big ) \Big | < \frac
C {|\tau|} ,$$

\noindent from which it follows that $|\eqref{eq:temp2}| < C
t^{-1}.$

To complete the estimate on \eqref{eq:j2} when $|\tau|/10 > M $,
it remains to bound

$$ \int_\R e^{-itF_\tau(\lambda)} \varphi (\lambda) \frac {2\lambda
}{ -\wp'(a)} (\chi(\lambda/N)-\chi(10 \lambda / |\tau|)-
\chi(\lambda + \tau /2) ) d\lambda  .$$

\noindent Integrating by parts, this equals

$$\frac 2{it} \int_\R e^{-itF_\tau(\lambda)}  \frac {d}{d\lambda}
\Big ( \frac {\lambda \varphi (\lambda)}{-\wp'(a)} \frac {
(\chi(\lambda/N)-\chi(10 \lambda/\tau)-\chi(\lambda + \tau /2)) }{
\partial_\lambda F_\tau (\lambda)} \Big )   d \lambda :=
 J_1 + J_2 ,
 $$

\noindent where

$$J_1 =  \frac 2{it}\int_\R e^{-itF_\tau(\lambda)} \frac {
(\chi(\lambda/N)-\chi(10 \lambda / |\tau|)-\chi(\lambda + \tau /2))
}{
\partial_\lambda F_\tau (\lambda)}  \frac
{d}{d\lambda} \frac {\lambda \varphi
(\lambda)}{-\wp'(a)}d\lambda,$$

and

$$J_2 =  \frac 2{it} \int_\R e^{-itF_\tau(\lambda)}\frac {\lambda \varphi (\lambda)}{-\wp'(a)}   \frac
{d}{d\lambda}  \frac { (\chi(\lambda/N)-\chi(10 \lambda /
|\tau|)-\chi(\lambda + \tau /2)) }{
\partial_\lambda F_\tau (\lambda)}  d \lambda .$$

In $J_2$, \be \partial_\lambda (\chi(\lambda/N)-\chi(10 \lambda /
|\tau|)-\chi(\lambda + \tau /2))= \frac 1N \chi'(\lambda/N) -
\frac {10}{|\tau|} \chi'(10 \lambda / |\tau|) - \chi'(\lambda +
\tau /2) .\label{eq:temp6}\ee

\noindent On the support of \eqref{eq:temp6}, we have $|\lambda
+\tau /2|
>1$ and $|\lambda| > M$. Thus $ |(\partial_\lambda F_\tau
(\lambda))^{-1}| <C$ by \eqref{eq:grow}. Consequently,

$$\Big |\frac 2{it} \int_\R e^{-itF_\tau(\lambda)}\frac
{\lambda \varphi (\lambda)}{-\wp'(a)  \partial_\lambda F_\tau
(\lambda)} \Big ( \frac 1N \chi'(\lambda/N) - \frac {10}{|\tau|}
\chi'(10 \lambda / |\tau|) - \chi'(\lambda + \tau /2) \Big ) d
\lambda \Big | < C t^{-1}. $$

On the support of $ \chi(\lambda/N)-\chi(10 \lambda /
|\tau|)-\chi(\lambda + \tau /2)$ we have $|\lambda | > | \tau /10|
$. It follows from \eqref{eq:symp3} that $ |\partial^2_\lambda
F_\tau (\lambda)| < 3$. Combining it with \eqref{eq:grow}, we obtain

$$\partial_\lambda  (\partial_\lambda F_\tau (\lambda))^{-1} < C (\lambda + \frac \tau 2)^{-2}, $$

\noindent which is integrable. Therefore

$$ \Big |\frac 2{it} \int_\R e^{-itF_\tau(\lambda)}\frac {\lambda \varphi (\lambda)}{-\wp'(a)}
\Big ( \partial_\lambda  \frac 1{ \partial_\lambda F_\tau (\lambda)}
\Big )
  (\chi(\lambda/N)-\chi(10 \lambda / |\tau|)-\chi(\lambda + \tau
/2))  d \lambda  \Big | < C t^{-1} .$$

\noindent This completes the estimate on $J_2$.

As for $J_1$, integrating by parts again, we obtain

$$J_1= - \frac 4{ t^2}\int_\R e^{-itF_\tau(\lambda)} \frac d
{d\lambda} \Big ( \frac { (\chi(\lambda/N)-\chi(10 \lambda /
|\tau|)-\chi(\lambda + \tau /2)) }{ (
\partial_\lambda F_\tau (\lambda) )^2 }  \frac
{d}{d\lambda} \frac {\lambda \varphi (\lambda)}{-\wp'(a)}  \Big ) d
\lambda .$$

\noindent Applying the Leibnitz's rule, we are left with three
terms. Two terms come from $\frac {d}{d \lambda}$ hitting $
\chi(\lambda/N)-\chi(10 \lambda / |\tau|)-\chi(\lambda + \tau /2)$
and $ (\partial_\lambda F_\tau (\lambda))^{-2}$, and the analysis
is analogous to that of $J_2$. When $\frac {d}{d \lambda}$ hits
 $ \frac {d}{d\lambda} \frac {\lambda \varphi
(\lambda)}{-\wp'(a)}$, we obtain the third term

$$- \frac 4{t^2} \int_\R e^{-itF_\tau(\lambda)} \frac {
\chi(\lambda/N)-\chi(10 \lambda / |\tau|)-\chi(\lambda + \tau /2) }{
(\partial_\lambda F_\tau (\lambda))^2}  \frac {d^2}{d\lambda^2}
\frac {\lambda \varphi (\lambda)}{-\wp'(a)}    d \lambda .$$

\noindent Because $|(\partial_\lambda F_\tau (\lambda))^2| >
|\lambda + \tau /2|^2/4$ on the support of
$\chi(\lambda/N)-\chi(10 \lambda / |\tau|)-\chi(\lambda + \tau
/2)$ and  $\frac {d^2}{d\lambda^2} \frac {\lambda \varphi
(\lambda)}{-\wp'(a)}$ is uniformly bounded, the above term is
dominated by $C t^{-2}$, where the constant $C$ is independent of
$N$.

This completes the estimate of \eqref{eq:j2} when $ |\tau| / {10}
>M$.  The analysis is similar and even simpler when $  |\tau | /10  \leq M$.
Therefore, when $P(x)$ has no double root in
$(-\infty,\wp(\omega_3)] $, we have

$$ \sup_{x,x'} |K_2(t,x,x')| < C t^{-\frac 13}.  $$

\noindent The decay factor $ t^{-\frac 13}$ is replaced by
$t^{-\frac 14}$ when $P(x)$ has a double root in
$(-\infty,\wp(\omega_3)] $.  $\Box$

\bigskip
Combining the estimates on $K_1(t,x,x')$ and $ K_2(t,x,x')$, we have
proved \eqref{eq:dec1} under the assumption \eqref{eq:simple}. We
have also proved \eqref{eq:dec2} for all nonzero $\omega , \omega '
\in \R$.

\bigskip

It remains to prove \eqref{eq:simple} holds for almost all
$\omega,\omega' \in \R$.

 Suppose $P(x)$ has a double root $x_0 \in
(-\infty, \wp(\omega_3) ] $. Then $x_0$ is a root of $P'(x)$. Recall
$$P'(x)= 6x^2 + \frac {12 \zeta(\omega)}{\omega} x + \frac {g_2}{2}
,$$

\noindent with its roots   $$r_+ , r_- = - \frac
{\zeta(\omega)}{\omega} \pm \sqrt{\Big ( \frac
{\zeta(\omega)}{\omega} \Big )^2 - \frac {g_2}{12}}\,\, .$$

That $P(x)$ has a double root in $ (-\infty, \wp(\omega_3) ] $
implies that $ (  \zeta(\omega)/\omega  )^2 - g_2/12 >0$ and $
P(r_-) =0 $. By \eqref{eq:gg}, $g_2$ and $g_3$ are real analytic for
$\omega, \omega ' \in \R^+$. By \eqref{eq:zeta}, $ \zeta(\omega) $
is also real analytic for $\omega, \omega' \in \R^+$. Therefore,
$r_+$, $r_-$ are analytic when $\omega, \omega ' \in \R^+$, with
branches at $\Big ( \frac {\zeta(\omega)}{\omega} \Big )^2 - \frac
{g_2}{12} =0 $. To prove \eqref{eq:simple} is true for almost all
$\omega, \omega ' \in \R^+$, it suffices to show that $ P(r_-)$ is
nonzero at one point. This can be done by direct numerical
calculation.

For example, take $\omega = 5.5$ and $\omega'=2$. Then we have

$$g_2= 0.507343, \qquad g_3= -0.0695438,$$

$$r_+, r_- = 0.0628169 \pm
0.195787 i, $$

$$P(r_+) , P(r_-) = -0.0386656 \pm 0.0300201 i .$$

This indicates that $ P(r_-) $ is nonzero for almost all $\omega,
\omega' \in \R$. Therefore, \eqref{eq:simple} holds for almost all
$\omega, \omega' \in \R$.

\section{Optimality of the decay factor }

So far we have proved the first part of Theorem~\ref{thm:main1}.
To verify \eqref{eq:opti}, we first reduce it to showing that
there exist constants $c>0$ and $T>0$ such that for  $t>T$

\be \|K(t,x,x') \|_{L^\infty} > c \, t^{-\frac 13}.
\label{eq:optimal} \ee

Accepting \eqref{eq:optimal} temporarily, we obtain that for any
given large $t$, there exist $(x_0, x'_0)$ such that $x_0 \neq x'_0$
and $|K(t,x_0,x'_0)|> c t^{-\frac 13}$. Without loss of generality,
suppose that

$$\Re (K(t,x_0,x'_0)) > \frac c2 t^{-\frac 13} .$$

 As $K(t,x,x')$
is smooth away from $x=x'$, there exists $\delta > 0$ such that
for any $ (x, x') \in (x_0- \delta, x_0+ \delta) \times (x'_0-
\delta, x'_0+ \delta) $

 $$\Re (K(t,x ,x' ) )  > \frac c4 t^{-\frac
13}.$$

\noindent Take the initial data $\psi_0(x')= \frac 1 {2\delta}
\chi_{(x'_0- \delta, x'_0+ \delta)}(x')$. Then $\|\psi_0 \|_{L^1} =1
$ and for any $x \in (x_0- \delta, x_0+ \delta) $

$$ |\psi(t,x)|= \Big |\int K(t,x,x')\psi_0(x')dx' \Big | >  \frac c4 t^{-\frac 13}. $$

To prove \eqref{eq:optimal}, we need the following lemma (Prop.3
Chap.\textrm{8} \cite{St}):

\begin{lemma}

Suppose $k \geq 2$, and $$\phi(x_0)=\phi'(x_0)= \cdots
=\phi^{(k-1)}(x_0)=0 ,$$ while $\phi^{(k)}(x_0) \neq 0 $. If $\psi
$ is supported on a sufficiently small neighborhood of $x_0$ and
$\psi(x_0) \neq 0 $, then

$$ \int_{\R} e^{i\lambda \phi(x)} \psi(x)dx = a_k  \psi(x_0) (\phi^{(k)}(x_0))^{-\frac 1k}
\lambda ^{-\frac 1k } + O(\lambda ^{-\frac 1k -1 } ),$$

\noindent where $ a_k \neq 0$ only depends on $ k$. The implicit
constant in $O(\lambda ^{-\frac 1k -1 } ) $ depends on only
finitely many derivatives of $\phi$ and $\psi$ at $x_0$.
\label{lemma:asmp}

\end{lemma}

By Lemma~\ref{lemma:root}, $P(a)$ has a unique simple root in
$(\wp(\omega_2), \wp(\omega_1)) $, thus we can choose $a_0
 \in (\omega_1, \omega_2 )$ and a corresponding $\tau_0$ such that
 both  $ \partial_a F_\tau(a)$ and  $ \partial^2_a F_\tau(a)$ vanish at $(a_0,\tau_0)$.

First, we denote $ I=[0,i2\omega']\cup [\omega_1,\omega_1+ i2\omega'
]$ and assume that for any $a \in I $, $a \neq a_0$, at least one of
$\partial_a F_{\tau_0}(a)$ and $
\partial^2_a F_{\tau_0}(a)$ does not vanish. Then we take $\delta
>0$ small enough such that for $a \notin ( a_0 - \delta,
 a_0+\delta) \subset I$, $ |\partial_a F_{\tau_0}(a)|+ | \partial^2_a F_{\tau_0}(a)|$ is
 greater than some positive constant.

Given any large $t$, take $(x,x')$ such that $ \frac {x-x'}{t}=
\tau_0 $ and

$$K(t,x,x')= \Big ( \int_{0}^{i 2 \omega'} + \int_{\omega_1}^{\omega_1+ i
2 \omega'} \Big ) e^{-itF_{\tau_0}(a)} m_a(x)m_{-a}(x')\frac
{-\wp'(a)da}{W(a)}.$$

\noindent The $\int_{0}^{i 2 \omega'}$-term is bounded by $C \,
t^{-\frac 12}$ using an argument analogous to that in Section 4,
because $|\partial_a F_{\tau_0}(a)|+ |
\partial^2_a F_{\tau_0}(a)| $ is uniformly greater than some
positive constant for $a \in (0, i 2 \omega')$.

We decompose the $\int_{\omega_1}^{\omega_1+ i 2 \omega'}$-term as
follows

$$\int_{\omega_1}^{\omega_1+ i 2 \omega'} e^{-itF_{\tau_0}(a)}
m_a(x)m_{-a}(x')\frac {-\wp'(a)da}{W(a)}:= J_3 +J_4 ,$$

\noindent where

$$  J_3 = \int_{\omega_1}^{\omega_1+ i 2 \omega'} e^{-itF_{\tau_0}(a)}
\rho(a)  m_a(x)m_{-a}(x')\frac {-\wp'(a)da}{W(a)} ,$$

\noindent and
$$
J_4= \int_{\omega_1}^{\omega_1+ i 2 \omega'} e^{-itF_{\tau_0}(a)}
 \tilde{\rho}(a) m_a(x)m_{-a}(x')\frac {-\wp'(a)da}{W(a)}.
$$

\noindent Here $\rho(a)$ is a smooth cut-off function supported on $
( a_0 - \delta, a_0+\delta)$ and $\tilde{\rho}(a)=1-\rho(a). $

Under our assumption, $|J_4| < C\, t^{-\frac 12}$, following the
same reasoning as that in Section $3$.

Considering $J_3$, the phase function $F_{\tau_0}(a)$ satisfies
$\partial_a  F_{\tau_0}(a_0)= \partial^2_a F_{\tau_0}(a_0)=0$ and
$\partial^3_a  F_{\tau_0}(a_0) \neq 0$. $m_a(x) $ and $m_{-a}(x')$
do not vanish when $a \in (\omega_1, \omega_2)$ by Eq.\eqref{eq:ma}.
$ \frac {-\wp'(a)}{W(a)} $ is nonzero when $a \in (\omega_1,
\omega_2)$. Therefore $m_{a_0}(x) m_{-a_0}(x') \frac
{-\wp'(a_0)}{W(a_0)} $ is nonzero.

 Since $\rho(a)$ is supported
in a sufficiently small neighborhood of $a_0$, by
Lemma~\ref{lemma:asmp}, there exist $c_1 >0$ and $T>0$ such that for
$t>T$

$$ |J_3| > c_1 t ^{-\frac 13},$$

\noindent where $c_1$ is independent of $t$.

Combining these estimates, we have $|K(t,x,x')|>  c_1 t ^{-\frac
13}- 2 C t ^{-\frac 12} > c_1 /2 \,\, t ^{-\frac 13} $ for any
$(x,x')$ satisfying $(x-x')/t= \tau_0 $ , which implies
\eqref{eq:optimal}.

Second, suppose there are other $a_1, \, a_2 \in I $ such that $a_1,
\, a_2, \, a_0$ are distinct and $\partial_a F_{\tau_0}(a)$, $
\partial^2_a F_{\tau_0}(a)$ both vanish at $a=a_1,\, a_2$. Then $P(x)$
 vanishes at $\wp(a_j)$, $j=0,1,2$.

Since $a_0 \in (\omega_1, \omega_2)$, we have that $-i \wp'(a_0) >0$
and $\zeta(\omega)/\omega + \wp(a_0) >0 $ by Lemma~\ref{lemma:root}.
Thus $\tau_0 <0$ by \eqref{eq:c1}. Similar analysis shows that when
$a \in (\omega_2, \omega_2 + i  \omega') \cup (  i  \omega',  2 i
\omega')$, $
\partial_a F_{\tau_0}(a) \neq 0$. Therefore, $a_1,\, a_2 \in (0, i\omega')$.

Thus $\wp(a_j)$, $j=0,1,2$ are distinct and are the three roots of
$P(x)$. This implies that there is no other $a \in I$ such that
$\partial_a  F_{\tau_0}(a)= \partial^2_a F_{\tau_0}(a)=0$.

We again set $\delta >0$ small enough such that for $a \notin
\bigcup_{j=0}^3 ( a_j - \delta,
 a_j+\delta) \subset I$, $ |\partial_a F_{\tau_0}(a)|+ | \partial^2_a F_{\tau_0}(a)|$ is
uniformly greater than some positive constant. Given any large $t$,
take $(x,x')$ such that $ \frac {x-x'}{t}= \tau_0 $. The earlier
argument implies

$$ K(t,x,x') = \sum_{j=0}^2 \int_{I} e^{-itF_{\tau_0}(a)}
\rho_j(a) m_a(x)m_{-a}(x')\frac {-\wp'(a)da}{W(a)} + O(t^{-\frac
12}) ,
$$

\noindent where $\rho_j(a)=1 $ when $|a-a_j| < \delta$ and
$\rho_j(a)=0 $ when $|a-a_j| > 2 \delta$.

\noindent By Lemma~\ref{lemma:asmp},

$$K(t,x,x') = a_3 t^{-\frac 13} \sum_{j=0}^2 (F_{\tau_0}^{(3)} (a_i) )^{-\frac 13}
 m_{a_j}(x)m_{-a_j}(x')\frac {-\wp'(a_j)}{W(a_j)} + O(t^{-\frac 12}).$$

\noindent Recall that $ x$ and $x'$ are related by $(x - x')/t =
\tau_0$. $m_{a_j}(x)m_{-a_j}(x')$, $j=0,1,2$, are linearly
independent as functions of $x \in \R$ and their nontrivial linear
combination is a nonzero function. Therefore, there exist $x_0$ and
$ x'_0$, satisfying $(x_0 - x'_0)/t = \tau_0$ and

$$K(t,x_0,x'_0) = c \,  t^{-\frac 13} +O(t^{-\frac 12}) ,$$

\noindent where $c$ is nonzero. Thus there exists $T$ such that
for $t>T$

$$ |K(t,x_0,x'_0)| > \frac c 2 \, t^{-\frac 13}. $$

Finally, suppose that $a_1=a_2$ in the second case, which is
equivalent to that $\wp(a_1)$ is a double root of $P(x)$ in
$(-\infty, \wp(\omega_3))$. Similarly, we have for $t>T$

\bea K(t,x,x')&=& \sum_{j=0}^1 \int_{I} e^{-itF_{\tau_0}(a)}
\rho_j(a) m_a(x)m_{-a}(x')\frac {-\wp'(a)da}{W(a)} + O(t^{-\frac
12}) \nn \\
&=&  a_3 t^{-\frac 13} (F_{\tau_0}^{(3)} (a_0) )^{-\frac 13}
 m_{a_0}(x)m_{-a_0}(x')\frac {-\wp'(a_0)}{W(a_0)} + \nn \\
&& \qquad\qquad\qquad
  a_4 t^{-\frac
14} (F_{\tau_1}^{(4)} (a_1) )^{-\frac 14}
 m_{a_1}(x)m_{-a_1}(x')\frac {-\wp'(a_1)}{W(a_1)} + O(t^{-\frac
12}) . \nn \eea

\noindent Therefore, there exists $(x_0,x_0')$ such that
$|K(t,x,x')| > c \, t^{-\frac 14} > c \, t^{-\frac 13} $. This
completes the proof of \eqref{eq:optimal}.

Our proof also gives the optimality of \eqref{eq:dec2} in the case
that $P(x)$ has a double root in $ (-\infty,\wp(\omega_3)] $.

By the proof of Lemma~\ref{lemma:root}, we have the following
corollary:

\begin{cor}
Suppose that $ (\zeta(\omega)/\omega)^2 \leq  g_2/12$. Then for
$t>1$,

\be
 \| U(t)\psi_0 \|_{L^\infty} < C t ^{-\frac 13} \| \psi_0
 \|_{L^1 (R)} .\nn
\ee

\end{cor}

Set $\omega =1$; it follows from Eq.\eqref{eq:gg} and
\eqref{eq:zeta} that $\frac {g_2}{12} - (\zeta(\omega)/\omega)^2 $,
as a function of $\omega' >0$, is analytic and $\omega' =0$ is its
essential singular point. Numerical experiment indicates that
$g_2/12 - (\zeta(\omega)/\omega)^2  \approx 0.966104 $ when $\omega
=1$ and $\omega' > 5$. When $\omega =1$ and $\omega' \to 0^+ $,
$g_2/12 - (\zeta(\omega)/\omega)^2 $ assumes each real number
infinitely many times.

\section{Appendix}

Here we list some elementary properties of Weierstrass functions
(\cite{WW}, \cite{Ch}, \cite{Ak}, \cite{GH}). A doubly-periodic
function which is meromorphic is called an elliptic function.
Suppose that $2\omega_1$ and $2\omega_3$ are two periods of an
elliptic function $f(z)$ and $\Im (\omega_3/\omega_1) \neq 0 $. Join
in succession the points $0,\, 2\omega_1,\,
2\omega_1+2\omega_3,\,2\omega_3,\,0 $ and we obtain a parallelogram.
If there is no point $\omega$ inside or on the boundary of this
parallelogram (the vertices excepted) such that $ f(z +
\omega)=f(z)$ for all values of $z$, this parallelogram is called a
fundamental period-parallelogram for an elliptic function with
periods $2\omega_1$ and $2\omega_3$. As a set, we assume this
parallelogram only includes one of four vertices and two edges
adjacent to it. In this way, the $z$-plane can be covered with the
translations of this parallelogram without any overlap. It can be
shown that for any $c \in \C$, the number of roots (counting
multiplicity) of the equation

 $$f(z)=c $$

\noindent  which lie in the fundamental period-parallelogram does
not depend on $c$. This number is called the order of the elliptic
function $f(z)$ and it equals the number of poles of $f$ inside a
fundamental period-parallelogram.

Given $\omega_1,\omega_3 \in \C$ with $\Im  (\omega_3/\omega_1)
\neq 0 $, the Weierstrass elliptic function is defined as

$$\wp(z) = \frac 1{z^2} + \sum_{(m,n)\neq (0,0)} \big \{
 (z-2m \omega_1 - 2n \omega_3)^{-2} - (2m \omega_1 +
2n \omega_3)^{-2} \big \}. $$

\noindent The summation extends over all integer values of $m$ and
$n$, simultaneous zero values of $m$ and $n$ excepted. $\wp(z)$ is
doubly-periodic, namely

$$\wp(z)= \wp(z + 2\omega_1) = \wp(z + 2\omega_3) .$$

$\wp(z)$ is an elliptic function of order $2$, with poles
$\Omega_{m,n}= 2m\omega_1 + 2n\omega_3$. Each pole $\Omega_{m,n}$
is of degree $2$. $\wp(z)$ is an even function, $\wp(z)=\wp(-z)$.
The Laurent's expansion of $\wp(z)$ at $z=0$ is written as

$$ \wp(z)= z^{-2} + \frac 1{20} g_2 z^2 + \frac 1{28}g_3 z^4 +
O(z^6),$$

\noindent where $g_2, g_3$ are the constants in Eq.\eqref{eq:dif1}
and \eqref{eq:dif2}. Explicitly, we have

\be g_2 = 60 \sum_{(m,\, n)\neq (0,0)} \Omega_{m,n}^{-4}, \quad \,\,
g_3 = 140 \sum_{(m,\, n)\neq (0,0)} \Omega_{m,n}^{-6}. \label{eq:gg}
\ee  Here $g_2$ and $g_3$ are called the invariants of $\wp$ and
they uniquely characterize $\wp$.

Since $\wp'$ is odd and elliptic of order $3$, it has three zeroes
in its fundamental period-parallelogram.  It is clear that these
zeroes are the half periods $\omega_1, \omega_2= \omega_1+ \omega_3
$ and $ \omega_3$. Denote $e_j = \wp(\omega_j), j=1,2,3$. The fact
that $\wp(z)$ is of order $2$ implies that $e_1,e_2,e_3$ are
distinct and that $\wp''$ does not vanish at $\omega_j$, $j=1,2,3$.
Furthermore, Eq.\eqref{eq:dif1} implies that $e_1,e_2,e_3$ are the
roots of the cubic polynomial

\be 4 x^3 -g_2 x -g_3=0 .  \label{eq:app3}\ee

The function $\zeta(z)$ is defined by the equation

$$ \frac {d}{dz} \zeta(z)= -\wp(z) ,$$

\noindent coupled with the condition $\lim _{z\to 0} (\zeta(z)
-z^{-1} )=0 $. $\zeta(z)$ may also be represented as

\be \zeta(z) = \frac 1z + \sum_{(m,\, n)\neq (0,0)} \Big \{ \frac
1{z-2m\omega_1 -2n\omega_3} + \frac 1{2m\omega_1 + 2n\omega_3}+\frac
z{(2m\omega_1 +2n\omega_3)^2} \Big \}. \label{eq:zeta} \ee

 \noindent $\zeta(z)$ is an odd meromorphic function of $z$ over the whole complex plane
except at simple poles $\Omega_{m,n}$. The residue at each pole is
$1$.

Write $\zeta(\omega_1) = \eta_1$ and $\zeta(\omega_3) = \eta_3 $;
then

$$\eta_1 \omega_3 -\eta_3 \omega_1 = \frac 12 \pi i .$$

$\zeta(z)$ is not doubly-periodic, however, it satisfies the
following equations

\be \zeta(z+ 2\omega_1)= \zeta(z)+ 2 \eta_1, \quad \quad  \zeta(z+
2\omega_3)= \zeta(z)+ 2 \eta_3. \label{eq:app1} \ee

Next we define $\sigma(z)$ by the equation

$$ \frac d{dz} \log \sigma (z) = \zeta (z), $$

\noindent coupled with the condition $\lim_{z \to 0} \sigma(z) / z =
1$. $\sigma(z)$ is an odd entire function with simple zeroes at
$\Omega_{m,n}$. Just like $\zeta(z)$,  $\sigma(z)$ satisfies

\be \sigma(z+ 2\omega_1)=- \sigma(z) e^{2\eta_1(z+\omega_1)},
\quad \quad
   \sigma(z+ 2\omega_3)= -\sigma(z)e^{2\eta_1(z+\omega_3)}. \label{eq:app2} \ee

If we assume that $\omega_1 =\omega , \,\, \omega_3 =i \omega' $ and
$\omega, \, \omega' \in \R $, then by symmetry $\wp(z) $ is real
valued when $\Re z \in \{ 0, \omega_1\} $ or $ \Im z \in \{ 0,
i\omega_3\}$. $\zeta(z)$ is real valued on the real line and is pure
imaginary when $\Re z=0$. Let $D$ to be the rectangle with vertices
$0, \omega , \omega+ i\omega'$ and $ i\omega'$. Then $\wp(z) $ sends
$D$ to the upper half plane conformally. As $z$ moves clockwise on
the boundary of $D$ both starting and ending at $0$, $\wp(z)$ varies
from $-\infty $ to $\infty$. This implies that $ \wp(  i\omega' )
<\wp( \omega+ i\omega' ) < \wp(  \omega)$.

\bibliographystyle{amsplain}

\noindent \textsc{Kaihua Cai: Division of Astronomy, Mathematics,
and Physics,
253-37 Caltech, Pasadena, C.A. 91125, U.S.A.}\\
{\em email: }\textsf{\bf kaihua@its.caltech.edu}

\end{document}